# Implementation of Tic-Tac-Toe Game in LabVIEW

**Lalitha Saroja Thota[1], Manal Elsayeed[1], Naseema Shaik[1], Tayf Abdullah Ghawa[1], Ahlam Rashed[1], Mona Refdan[1], Wejdan Mohammed[1], Rawan Ali[1]**
**Suresh Babu Changalasetty[2]**

[1] *Department of Computer Science, College of Arts and Science-1, Khamees Mushayat*
*King Khalid University, Abha, Kingdom of Saudi Arabia*
[2] *Department of Computer Engineering, College of Computer Science*
*King Khalid University, Abha, Kingdom of Saudi Arabia*

***ABSTRACT :*** *Tic-Tac-Toe game can be played by two players where the square block (3 x 3) can be filled with a cross (X) or a circle (O). The game will toggle between the players by giving the chance for each player to mark their move. When one of the players make a combination of 3 same markers in a horizontal, vertical or diagonal line the program will display which player has won, whether X or O. In this paper, we implement a 3x3 tic-tac-toe game in LabVIEW. The game is designed so that two players can play tic-tac-toe using LabVIEW software. The program will contain a display function and a select function to place the symbol as well as toggle between the symbols allowing each player a turn to play the game. The program will update after each player makes their move and check for the conditions of game as it goes on. Overall program works without any bugs and is able to use*

## I. INTRODUCTION

Games provide a real source of enjoyment in daily life. Games also are helpful in improving the physical and mental health of human. Apart from daily life physical games, people also play computer games. These games are different than those of physical games in a sense that they do not involve much physical activity rather mental and emotional activities. Getting games to react back to the user of a game has always been long hard question for game programmers. Because, lets just face it, a good game that doesn't challenge the user's ability to play the game doesn't keep the user around very long. This idea can be applied to any form of game that is out there. Board games are never fun when the opponent that he or she is playing doesn't learn or catches on. With today's computers always advancing, programmers are always looking for new ways to make a video game more interesting and challenging for the user.

Tic-Tac-Toe game [1] can be played by two players where the square block (3 x 3) can be filled with a cross (X) or a circle (O). The game will toggle between the players by giving the chance for each player to mark their move. When one of the players make a combination of 3 same markers in a horizontal, vertical or diagonal line the program will display which player has won, whether X or O.

The Tic-Tac-Toe game is most familiar among all the age groups. The friendliness of Tic-tac-toe games makes them ideal as a pedagogical tool for teaching the concepts of good sportsmanship. The game is a very good brain exercise. It involves looking ahead and trying to figure out what the person playing against you might do next.

## II. EXISTING SYSTEM

Tic-Tac-Toe (or Noughts and crosses, Xs and Os) is a pencil-and-paper game for two players, O and X, who take turns marking the spaces in a 3 x 3 grid. The player who succeeds in placing three respective marks in a horizontal, vertical or diagonal row wins the game.

## III. PROPOSED SYSTEM

In this paper, we create a 3x3 tic-tac-toe game in LabVIEW. The system is designed so that two players can play a game of tic-tac-toe using LabVIEW software [2]. The program will contain a display function and a select function to place the symbol as well as toggle between the symbols allowing each player a turn to play the game. The program will update after each player makes their move and will check for the conditions of the game as it goes on.

## IV. FEATURES OF SYSTEM

The computer implementation of the game Tic-Tac-Toe has many features as compared to the traditional way of playing it with paper and pencil. The various features are:
- The game has been made user friendly with proper use of LabVIEW software
- The user can play as many games without any interruption
- The user can choose any symbol he/she wants to
- The game has been made as a thorough expert system
- The player can win the game, draw the game or will loose the game
- It's a good brain exercise for all age group people

## V. RELATED WORK

We study several research papers on tic-tac-toe game and summarize the findings below.

Tic-Tac-Toe is a simple and yet an interesting board game. Researchers have used various approaches to study the Tic-Tac-Toe game. For example, Fok and Ong [3] and Grim et al. [4] have used artificial neural network based strategies to play it. Citrenbaum [5] and Yakowitz [6] discuss games like Go-Moku, Hex and Bridg-It which share some similarities with Tic-Tac-Toe.

Many software implementations of Tic Tac game had been reported and recently it became available for smart phone such as the one for Apple iPhone [7], and other for Android environment [8].

Traditionally, the game of Tic-tac-toe is a pencil and paper game played by two people who take turn to place their pieces on a 3 times 3 grid with the objective of being the first player to fill a horizontal, vertical, or diagonal row with their pieces. What if instead of having one person playing against another, one person plays against a team of nine players, each of whom is responsible for one cell in the 3 times 3 grid? In this new way of playing the game, the team has to coordinate its players, who are acting independently based on their limited information. Soedarmadji [13] present a solution that can be extended to the case where two such teams play against each other, and also to other board games. Essentially, the solution uses a decentralized decision making, which at first seems to complicate the solution.

Stephen Mann and Matthew Netsch [9] design of a parallel digital circuit that performs neural network (NN) calculations to evaluate Tic-Tac-Toe position was introduced. FPGA's are programmed to implement custom digital designs by physically mapping paths between the logic gates on each device. Using an FPGA allows the structure of the NN to be reprogrammed without any monetary cost. The author claims that NN implementation has better performance than traditional software implementations, because it takes advantage of the NN's inherent parallel structure.

Another NN application is implemented by Shahzeb Siddiqui et al [10] extend the game by adding two additional rows, two additional columns, and has been extended to the 3rd dimension. The paper calculate the optimum position using the idea of creating a neural network that uses backpropagation coupled with elements of a genetic algorithm to improve the likelihood that the most optimal solution is obtained and outline our methodology at the implementation level.

Pinaki Chakraborty [11] formally defines the Tic-Tac-Toe game and then develops artificial intelligence based strategies to play the same. Leaw and Cheong [12] perform a minimalistic quantization of the classical game of tic-tac-toe, by allowing superpositions of classical moves.

An optically routed gate array (OPGA) is used by *Edward* [14] to implement a simple game of Tic-tact-toe to demonstrate the utility of electro-optical circuits which embody user input, display and logic functions in a single device. Progresses on routing techniques layout and logic simulation are also presented.

The hardware implementation of intelligent Tic-Tac toy is presented by Alauddin [15]. The implementation uses Graphical LCD (GLCD) touch screen and microcontroller. The microcontroller receives the player move from GLCD (displayed as X) and uses intelligent algorithm to analyze the move and choose the best counter move. The microcontroller displays the counter move on the screen as circle (O). The algorithm decides the winner when game is finished according to the Tic-Tac playing rule. The system is implemented using cheap available off-the-shelf electronic components and tested and proved to be working fast and efficiently

## VI. SYSTEM REQUIREMENTS

(a) Computer system with minimum P4 processor
(b) LabVIEW 2010 Software

## VII. SYSTEM DESIGN

The Use Case diagram for Tic-tac-toe game is shown in Fig 1

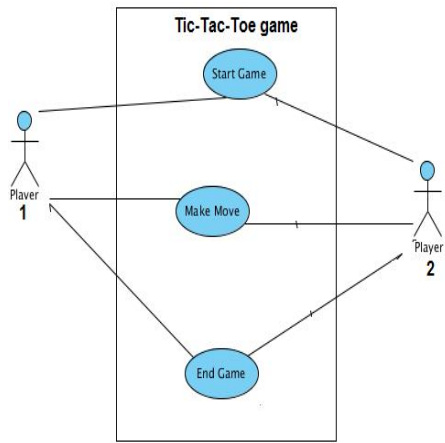

**Fig 1: Use Case diagram for Tic-tac-toe game**

The flow chart of tic-tac-toe game use in present system is shown in Fig 2

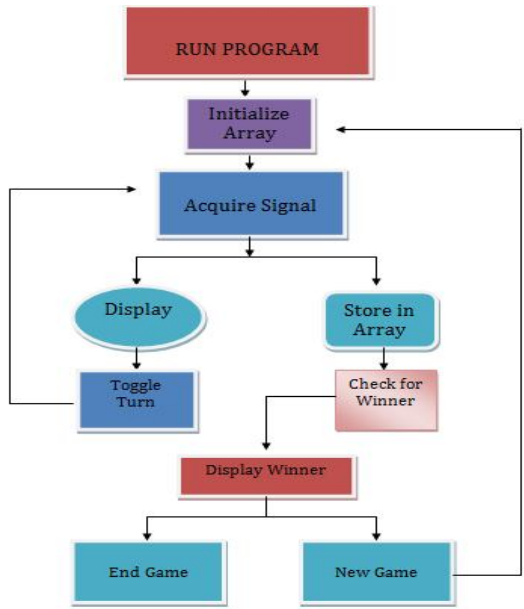

**Fig 2: Tic-tac-toe program flow chart**

*How to play Tic-Tac-Toe game?*
- The game is played between two players
- Both players choose their respective symbols to mark
- Player 1 starts the turn by placing his/her symbol on any of the nine squares
- Then player 2 marks the turn by placing his/her symbol on the empty squares
- Both players make their turns alternately
- If any player gets the three respective symbols in a horizontal, vertical or diagonal row wins the game

*Performing the Winner Checks*

In a tic-tac-toe game there are 8 possibilities to win the game as discussed above. The possibilities are getting the three combinations horizontally, three combinations vertically or two combinations diagonally as shown in the Fig 3. Therefore the easiest way to find the way to determine the winner of the game is to check for these eight combinations as the game goes on.

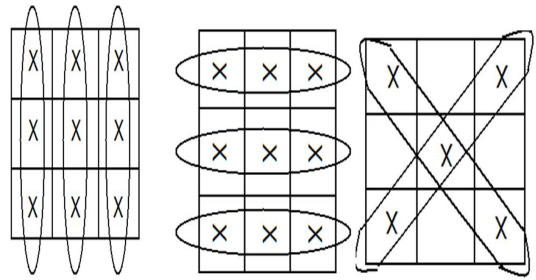

**Fig 3:Winning possibilities of Tic-Tac-Toe game**

The system will be designed so that two players can play a game of tic-tac-toe using LabVIEW software. The program will contain a display function and a select function to place the symbol as well as toggle between the symbols allowing each player a turn to play the game. Fig 4 shows the tic-tac-toe game interface.

Using LabVIEW the tic-tac-toe program will meet the following criteria:
- Store symbol value in an array to be displayed.
- Change player turn
- Rest game when clicked on new game
- Perform the checking of array symbols to determine a winner
- Display the winner of the current game

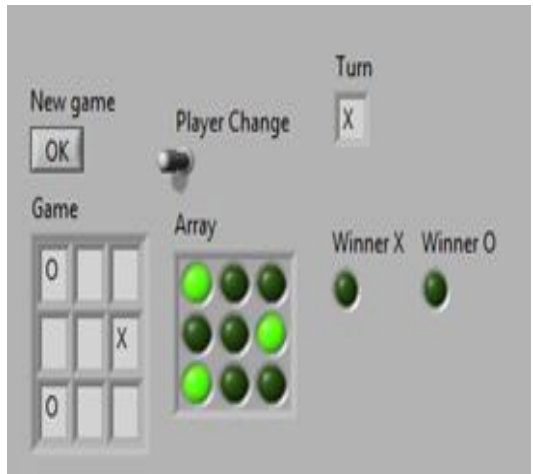

**Fig 4: Tic-tac-toe game interface**

## VIII. SYSTEM IMPLEMENTATION

Software implementation is the stage in the software engineering process at which an executable software system is developed. In this stage where the theoretical design is turned into a working system.

LabVIEW [2] – ( Laboratory Virtual Instrument Engineering Workbench) is developed by National Instruments. LabVIEW is graphical programming software that allows for instrument control, data acquisition, and processing of acquired data.

LabVIEW provides the flexibility of a powerful programming language without the complexity of traditional development environments. Some benefits of LabVIEW are
- Easy to Learn and Use
- Complete Functionality
- Integrated I/O Capabilities
- no need to write lines of program code
- Takes less time to develop software
- Cost and manpower required is minimum

LabVIEW programs are called virtual Instruments (VI). Stress that controls equal inputs, indicators equal outputs.

Each VI contains three main parts:
- Front Panel – How the user interacts with the VI.
- Block Diagram – the code that controls the program.
- Icon/Connector – Means of connecting a VI to other Vis

In order to acquire the players symbol and to display the symbols two array functions were created, one for displaying and another for acquiring since LabVIEW does not have one function to read and display at the same time. Once both the arrays were created they were set to 2 dimensional arrays. Fig 5 and 6 shows the creation of selection array and display array in LabVIEW.

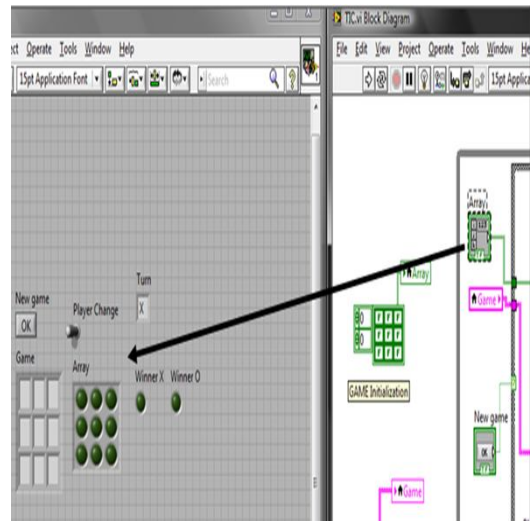

**Fig 5: Creation of a selection array**

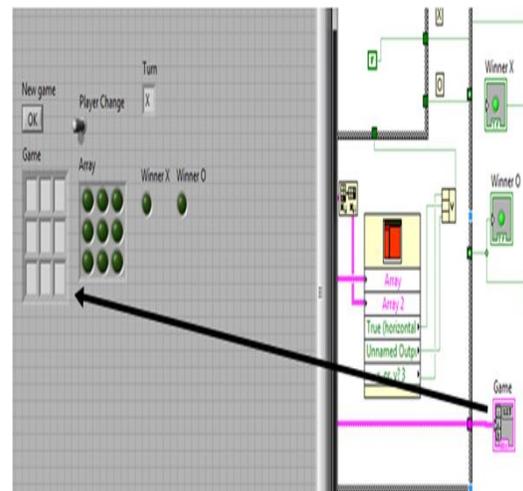

**Fig 6: Display array**

A sub VI that checks the horizontal, vertical and diagonal lines whether they have obtained three equal signs by using Boolean programming functions is shown in Fig 7. This function will check the condition each time a player has input his symbol. When the symbol is captured by the array then it will be passed into the sub VI where it will be transferred to one or zero as X(cross) will be one and O(nought) will be represented by zero. The sub VI will be checked vertically by using the transposed array function

Fig 10 show a part of VI where a case structure is used to select the winner. Fig 11 presents the total VI of the tic-tac-toe game

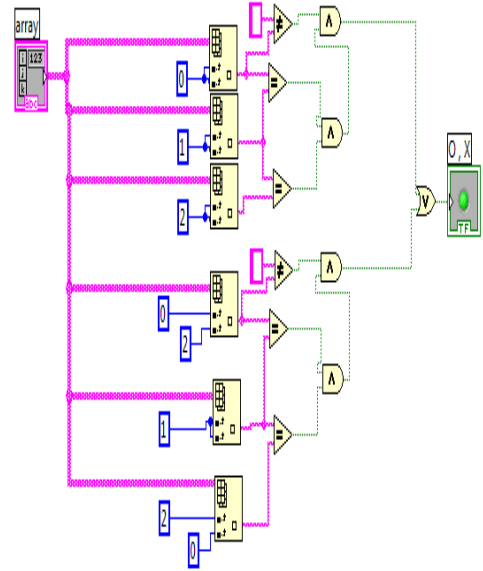

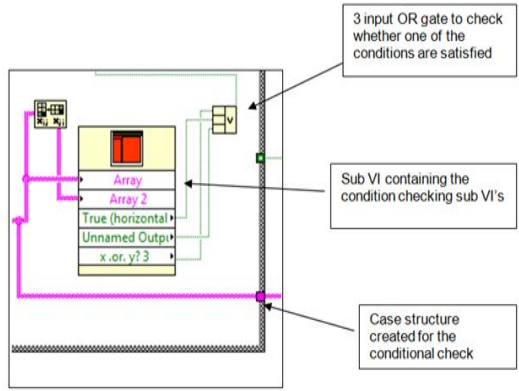

**Fig 7: Sub VI created for Combination check**

In the Fig 7 the array (Array) will be passes in to the sub VI for condition check to figure out a winner as well as the vertical possibility condition check array (Array 2). And output of each condition checking sub VI has been connected to a three input OR gate where it has been connected to a case structure to display the winner when the OR gate becomes TRUE.

Fig 8 and 9 shows each sub VI where the programming logic functions have been used to check each possibilities to check horizontal, diagonal lines of game.

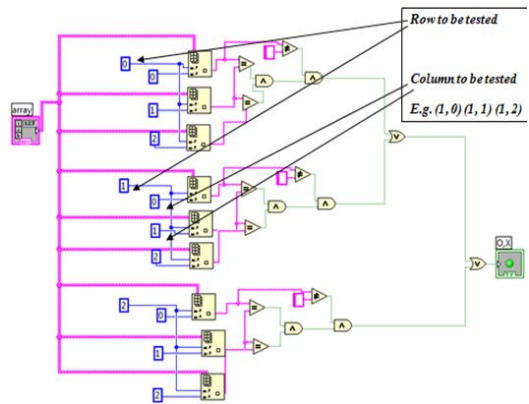

**Fig 8: Sub VI created to check horizontal lines of game**

**Fig 9: Sub vi created to check diagonal lines of game**

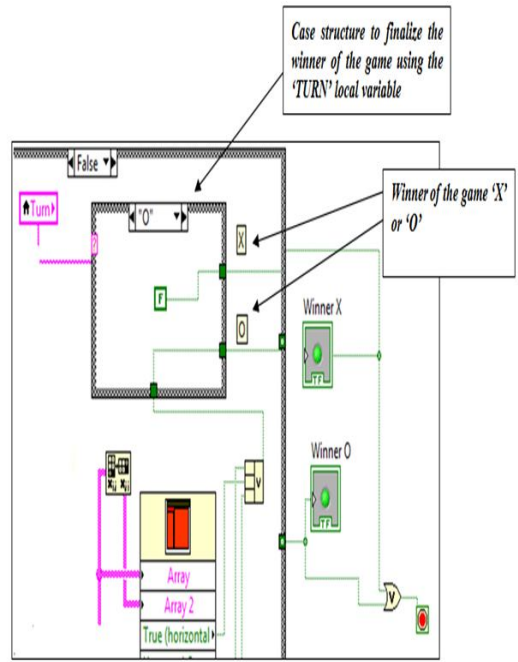

**Fig 10: Case structure to select a winner**

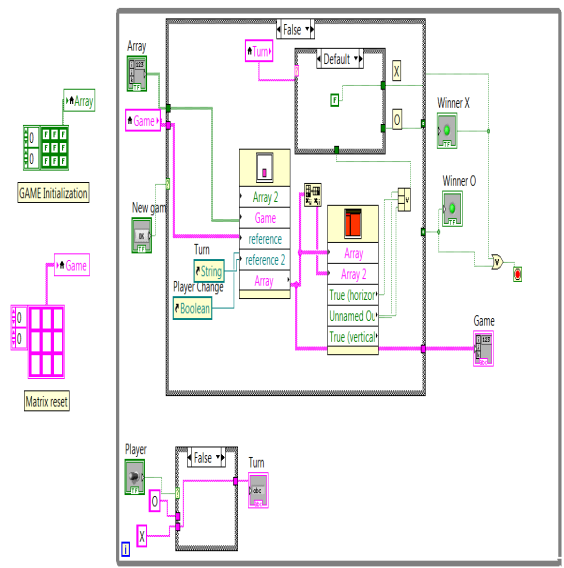

**Fig 11: Tic-tac-toe game Program VI**

## IX. SYSTEM TESTING

Testing is a process, which reveals errors in the program. Once the source code has been generated, the software must be tested to uncover as many errors as possible before delivery to the customer. Software testing can be stated as the process of validating and verifying that a computer program/application/product:

(a) meets the requirements that guided its design and development,
(b) works as expected,
(c) can be implemented with the same characteristics,
(d) satisfies the needs of stakeholders.

LabVIEW programs are called Virtual Instruments (VI). For the tic-tac-toe system various VIs are made, each VI represent one module in the project. Unit test is done for each VI. The errors are detected in LabVIEW while VIs are made. VI will not work if there are errors. This error needs to be rectified before a VI is run. All VIs are individually tested in unit testing and are successful.

In the system various VIs and sub Vis are linked together to make complete system. The integration of all VIs are tested in integrated testing and found successful. The links are found to work properly. The system is finally run for system testing to determine the system meets all users requirements as a whole and found successful.

## X. RESULTS AND DISCUSSIONS

The tic-tac-toe game is run in LabVIEW with two players. Fig 12 to 17 shows the screen shorts of game at various stages presenting player X choosing symbol, player O choosing symbol, Winner X, Winner O and draw game

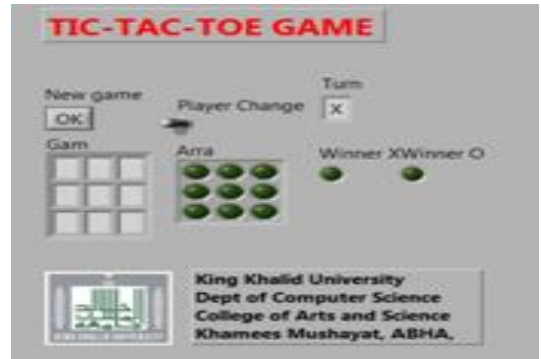

**Fig 12: Start of tic-tac-toe game, display turn of player X**

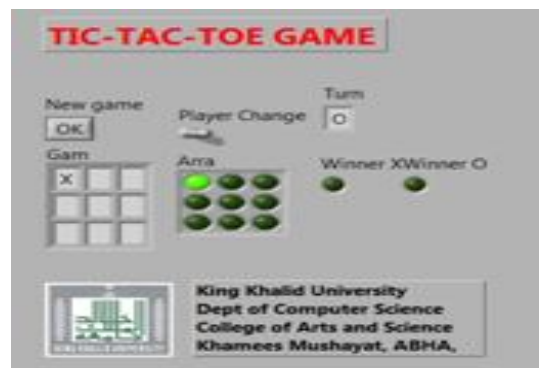

**Fig 13: Player X choose symbol and game toggles to player O, show turn O**

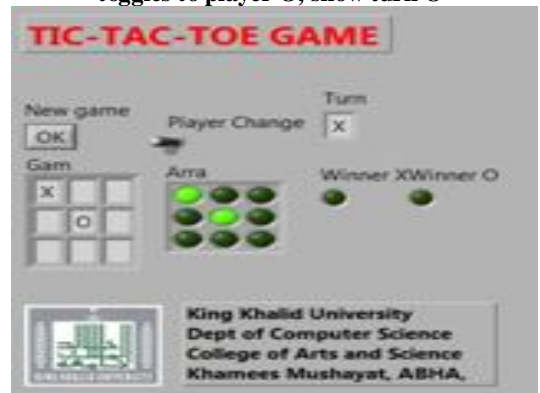

**Fig 14: Player O choose symbol and game toggles to player X, show turn X**

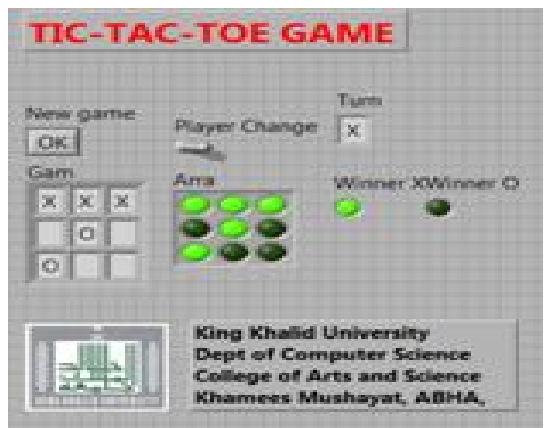

**Fig 15: Player X gets 3 symbols horizontal; game displays Winner X and ends**

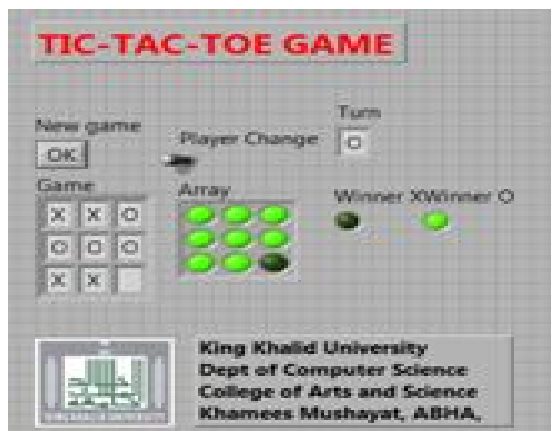

**Fig 16: Player O gets 3 symbols horizontal; game displays Winner O and ends**

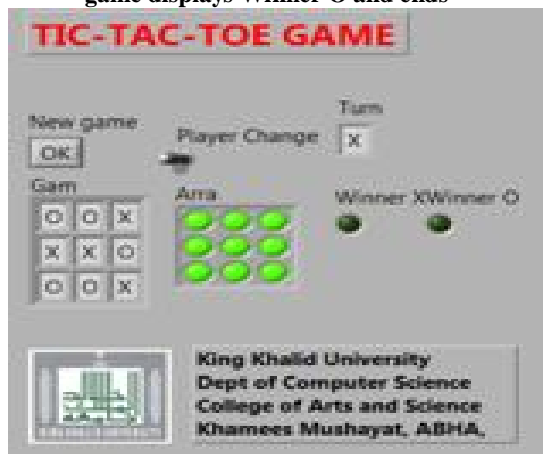

**Fig 17: No Player gets 3 symbols, game draw and asks for new game**

## XI. CONCLUSION

We develop and implement tic-tac-toe game in an event driven GUI software using the platform of the LabVIEW. In this system, we create a 3x3 tic-tac-toe game in LabVIEW. The system is designed so that two players can play a game of tic-tac-toe using LabVIEW software. The program will contain a display function and a select function to place the symbol as well as toggle between the symbols allowing each player a turn to play the game. The program will update after each player makes their move and will check for the conditions of the game as it goes on. Overall the system works without any bugs and is able to use.